# Enhancing Air Quality Monitoring: A Brief Review of Federated Learning Advances

Sara Yarham[1], Mehran Behjati[1], Haider A. H. Alobaidy[2], Anwar P.P. Abdul Majeed[1], Yufan Zheng[3]

[1] Department of Computing and Information Systems, School of Engineering and Technology, Sunway University, 47500, Malaysia
[2] Department of Information and Communications Engineering, College of Information Engineering, Al-Nahrain University, Baghdad 64046, Iraq
[3] School of Intelligent Manufacturing Ecosystem, XJTLU Entrepreneur College, Xi'an Jiaotong– Liverpool University (XJTLU), Taicang, Suzhou, 215400, China
mehranb@sunway.edu.my

**Abstract.** Monitoring air quality and environmental conditions is crucial for public health and effective urban planning. Current environmental monitoring approaches often rely on centralized data collection and processing, which pose significant privacy, security, and scalability challenges. Federated Learning (FL) offers a promising solution to these limitations by enabling collaborative model training across multiple devices without sharing raw data. This decentralized approach addresses privacy concerns while still leveraging distributed data sources.

This paper provides a comprehensive review of FL applications in air quality and environmental monitoring, emphasizing its effectiveness in predicting pollutants and managing environmental data. However, the paper also identifies key limitations of FL when applied in this domain, including challenges such as communication overhead, infrastructure demands, generalizability issues, computational complexity, and security vulnerabilities. For instance, communication overhead, caused by the frequent exchange of model updates between local devices and central servers, is a notable challenge. To address this, future research should focus on optimizing communication protocols and reducing the frequency of updates to lessen the burden on network resources. Additionally, the paper suggests further research directions to refine FL frameworks and enhance their applicability in real-world environmental monitoring scenarios.

By synthesizing findings from existing studies, this paper highlights the potential of FL to improve air quality management while maintaining data privacy and security, and it provides valuable insights for future developments in the field.

# 1  Introduction

Air quality monitoring has become increasingly critical as urbanization accelerates and the detrimental effects of pollution on human health and the environment become more apparent. The quality of the air we breathe has a significant impact on economic stability and public health. Accurate monitoring and management are necessary for environmental protection and regulatory compliance. A wide range of both artificial and natural causes influence air quality, including forest fires, climate change, ozone depletion, industrialization, urbanization, and transportation emissions. Sulfur dioxide ($SO_2$), Nitrogen dioxide ($NO_2$), Carbon dioxide ($CO_2$), Carbon monoxide (CO), nitrogen oxides (NOx), and particulate matter (PM2.5, PM10) are only a few of the many pollutants found in the atmosphere that can harm the environment and human health.

Significant research efforts have been dedicated to forecasting air pollution and predicting the Air Quality Index (AQI) on a global scale, focusing on pollutant forecasting. However, many of these studies have relied on traditional machine learning (ML) methods [1, 2], which involve centralizing data for processing. These centralized approaches pose several challenges, including data privacy concerns, high energy consumption, and the complexities of managing and processing large datasets.

Federated Learning (FL) offers a promising alternative by enabling collaborative learning across multiple devices without the need to centralize data. Introduced by Google in 2016 [3], FL allows devices to train models locally and then share only the model updates, thus preserving data privacy and reducing the need for large-scale data transfers. This approach not only addresses the privacy and security issues inherent in centralized ML but also improves model accuracy, response time, and adaptability to changing environmental conditions [4].

Considering the crucial role of air quality monitoring in mitigating air pollution, an issue that poses significant health risks [5] and contributes to climate change [6], there is a compelling need to develop and implement advanced techniques for accurate monitoring and forecasting of air quality.

Although previous research has explored the application of FL in air quality monitoring, these studies have not fully addressed the limitations associated with the current approaches [4, 7]. This paper seeks to address these gaps by critically evaluating existing FL applications in air quality monitoring and proposing directions for future research.

The research scope focuses on applying FL to air quality and environmental monitoring, particularly in predicting pollutants and improving data management. The approach involves a critical review of existing studies, identifying gaps, and proposing solutions to improve the effectiveness of FL in real-world scenarios. The scope includes an analysis of the practical challenges and opportunities presented by FL, with an emphasis on its potential to address the limitations of traditional ML methods. This review aims to contribute to the ongoing development of FL as a powerful tool for air quality monitoring and management. It highlights the present status of FL applications as well as the possibility of enhancing environmental monitoring through decentralized learning frameworks, providing useful insights for future developments.

The key contributions of this paper include:

1. Comprehensive Review of FL Applications in Air Quality Monitoring: An extensive review is conducted on how FL has been applied to air quality monitoring, focusing on predicting specific air pollutants such as PM2.5, PM10, $NO_2$, $SO_2$, CO, and O3. The review assesses the effectiveness of FL models in capturing complex spatial and temporal dependencies within air quality data while ensuring data privacy and security.
2. Identification of Limitations and Future Research Directions: The paper examines the limitations of current FL applications in air quality monitoring, including challenges such as communication overhead, computational complexity, scalability, generalizability, and security vulnerabilities. Potential future research directions are also outlined, including developing more efficient algorithms, enhanced security measures, and exploring FL's scalability across diverse environmental settings.

The paper is structured as follows: Section 2 provides a detailed overview of FL, including its training process and key categories. Section 3 reviews the current applications of FL in air quality monitoring, highlighting specific case studies and their outcomes. Section 4 discusses the limitations of FL when applied to this domain, including communication overhead, infrastructure requirements, generalizability, and security issues. Section 5 offers suggestions for future research directions aimed at addressing these challenges. Finally, Section 6 concludes the paper by summarizing the key findings and outlining the implications for future work in this field.

## 2   Federated Learning: Overview and Categories

In conventional machine learning (ML), training data is typically collected and stored on a central server, where the model is trained. This centralized approach can lead to significant privacy risks, as sensitive user data is aggregated and managed centrally, making it vulnerable to breaches and misuse [8].

FL offers a solution to these challenges by enabling multiple devices or nodes to collaboratively train a shared global model without the need to centralize data.

The typical FL training process involves several iterative steps, as illustrated in Fig.1. First, the central server selects a subset of devices, known as clients, to participate in the current training round. Once selected, the server broadcasts the current global model parameters to these clients. Each client then independently trains the model locally on its data, using the received parameters, and computes updates to the model. These updates are subsequently sent back to the central server. The server then aggregates these updates, often by averaging, to form an updated global model. This updated model is then sent back to the clients for the next round of training. This process is repeated until the model reaches the desired level of accuracy and performance [9].

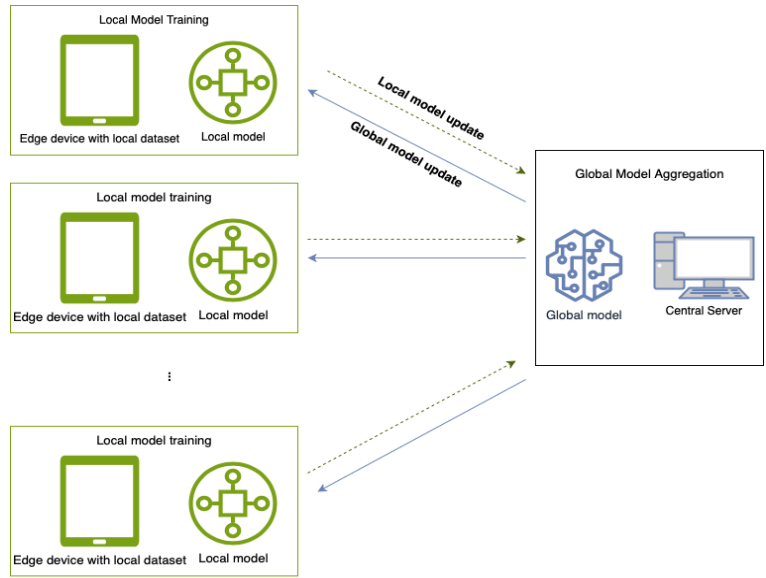

**Fig. 1.** FL training process.

FL is categorized into three primary categories [10], depending on how data is distributed across the clients and how the data is structured.

**Horizontal Federated Learning (HFL)** is applicable when different organizations or devices have datasets with the same feature space but different samples [11]. For example, multiple hospitals may collaborate using HFL, where each hospital has data on different patients but with the same medical features. The term "horizontal" refers to the distribution of data samples across clients. HFL is designed to handle scenarios where data is distributed so that each client has similar features but different instances or records. This is illustrated in Fig. 2.

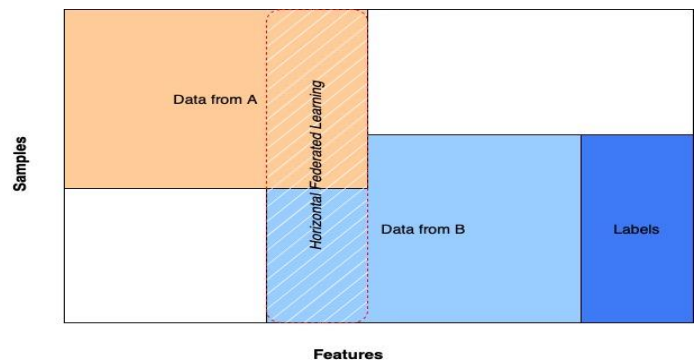

**Fig. 2.** HFL with common features and different samples [7].

**Vertical Federated Learning (VFL)** is used when different entities have datasets that share the same samples but have different features. For instance, a bank and an insurance company might both have data on the same customers but with different types of information. VFL enables these entities to train a model collaboratively without sharing their proprietary data. The term "vertical" indicates that data features are divided among different parties, but the entities (samples) remain consistent across datasets. This is illustrated in Fig. 3.

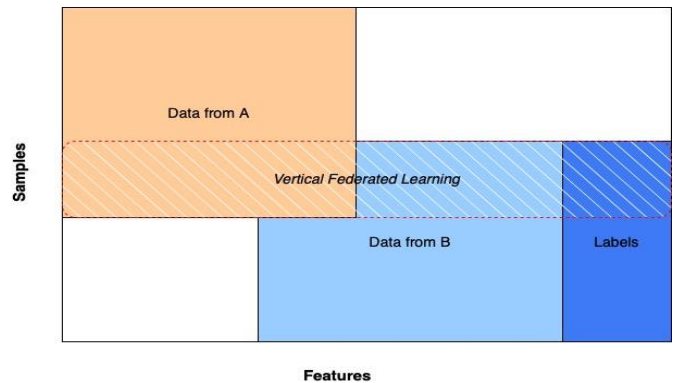

**Fig. 3.** VFL with common samples and different features [7].

**Federated Transfer Learning (FTL)** is employed when the datasets across different entities differ in both the sample space and the feature space. This category of FL leverages transfer learning techniques to enable collaboration between parties that have limited data overlap, either in terms of samples or features. FTL is useful in scenarios where the parties involved have different kinds of data and there is minimal direct overlap in the data they possess. This is illustrated in Fig. 4.

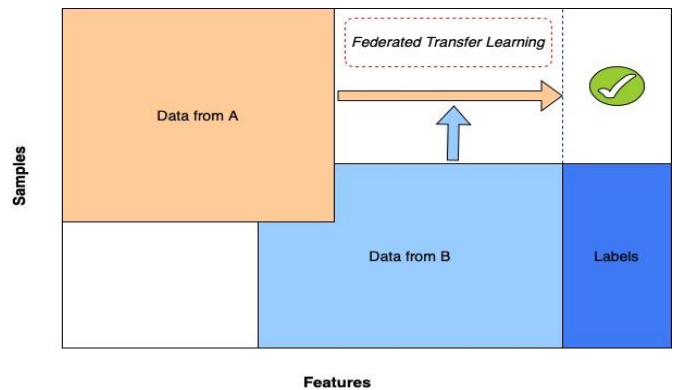

**Fig. 4.** FTL with distinct samples and features [7].

## 3 Application of Federated Learning in Air Quality Monitoring

### 3.1 Federated Learning for Predicting Air Pollutant Levels

Air pollutants such as particulate matter (PM2.5, PM10), NO2, SO2, CO, and O3 are critical indicators of air quality, directly impacting public health. Several studies have focused on applying FL to enhance the prediction of these pollutants.

Hu et al. [12] introduced FedDeep, a federated deep learning model designed to predict PM2.5 levels across urban regions in Jiangsu Province, China. By leveraging an External Spatio-Temporal Network (ESTNet), the model effectively captured spatial and temporal dependencies in air quality data, significantly outperforming traditional models. The FedDeep model achieved a Mean Absolute Error (MAE) of 12.38, a Root Mean Squared Error (RMSE) of 19.19, and a coefficient of determination ($R^2$) of 0.972 for 12-hour PM2.5 forecasting. Additionally, the model demonstrated efficiency in terms of reduced training time and GPU memory usage, with a total training time of 28.32 minutes and a GPU memory usage of 0.87 GB. Despite these successes, the study also highlighted challenges such as communication overhead and the need for robust network infrastructure to support model updates.

Similarly, Huang et al. [13] proposed a cross-domain prediction model that combines FL with the differential privacy Laplace mechanism (DPLA) and an optimized long short-term memory (LSTM) neural network using the Sparrow Search Algorithm (SSA). This model was applied to predict concentrations of multiple air pollutants, including PM2.5, PM10, SO2, NO2, O3, and CO across cities in China. The results demonstrated superior accuracy, with an RMSE of 10.59, an MAE of 6.26, and an $R^2$ of 92.93%, significantly outperforming baseline models such as multilayer perceptron (MLP), support vector regression (SVR), and traditional LSTM models. However, the study also identified challenges related to the model's generalizability, scalability, and potential vulnerabilities to adversarial attacks.

In another study, Abimannan et al. [14] developed a hybrid convolutional neural network-LSTM (CNN-LSTM) model within an FL framework to predict PM2.5 concentrations in Mumbai, India. This model effectively captured both spatial and temporal dependencies, achieving high prediction accuracy with an MAE of 0.466, an RMSE of 0.522, and an $R^2$ of 0.9877. However, the study's focus on a single geographic area raised concerns about its generalizability to other regions and conditions.

### 3.2 Federated Learning for Air Quality Index Prediction

AQI is a comprehensive metric that provides an overall assessment of air quality based on the concentrations of multiple pollutants. FL has been utilized in several studies to improve the accuracy and reliability of AQI predictions.

Chhikara et al. [15] introduced a decentralized FL framework within a swarm of unmanned aerial vehicles (UAVs) to predict AQI in Delhi, India. The study utilized UAVs to collect air quality data from various altitudes and locations, aiming to enhance the accuracy of AQI predictions. The proposed LSTM model within this framework demonstrated superior performance compared to traditional models, achieving an RMSE of 56.222, an MAE of 41.219, and a mean absolute percentage error (MAPE) of 24.184. Despite these promising results, the study faced challenges such as the need for frequent sensor recalibration, potential connectivity issues with UAVs, and limitations in the generalizability of the findings.

Dey and Pal [16] focused on improving AQI predictions using a Bidirectional Gated Recurrent Unit (BGRU) model within an FL framework, specifically applied to Indian smart cities. The model effectively handled long-term dependencies in AQI data and outperformed other machine learning models like support vector machine (SVM) and random forest (RF). The FL-BGRU model achieved an MSE of 40.129 and an MAE of 36.659, demonstrating its superiority over traditional models. However, the study raised concerns about the scalability of the model across a larger number of nodes (cities) and the generalizability of the results beyond the specific datasets used.

Jin et al. [17] proposed a Nested LSTM (NLSTM) network for predicting multiple AQI components, including PM2.5 and PM10, within a multi-task multi-channel FL framework. This approach allowed the model to leverage the internal correlations between different pollutants, achieving high prediction accuracy. The model achieved MAE values of 0.88 for PM2.5, 2.98 for PM10, and an $R^2$ value close to 0.99 for both pollutants. However, the model's complexity led to longer training times and raised concerns about its scalability to different geographic regions.

### 3.3 Federated Learning for Carbon Emission Prediction

While most studies have focused on air pollutants, some have explored the application of FL in predicting carbon emissions, which is a critical factor in managing industrial impacts on the environment and addressing climate change.

Cui et al. [18] combined FL with Seasonal AutoRegressive Integrated Moving Average (SARIMA) models to predict carbon emissions in the electricity sector across 13 countries. The FL approach, combined with SARIMA-based clustering, improved prediction accuracy and computational efficiency. The model showed significant reductions in MSE and MAE, particularly within optimally clustered clients, achieving an MSE improvement of 79.27% and an MAE improvement of 63.32%. Despite these improvements, the study's sector-specific focus limited its applicability to other industries, and the challenges in ensuring equitable benefits for all participants in the FL process were also highlighted.

### 3.4 Federated Learning for Data Imputation and Management

Ensuring the continuity and reliability of air quality data is critical for effective environmental management. FL has been applied to develop methods for imputing missing data, maintaining data integrity across distributed datasets.

Zhou et al. [19] introduced a Federated Conditional Generative Adversarial Nets (FCGAN) framework to address the issue of missing air quality data. By leveraging the FL framework, the model was able to generate accurate imputations while preserving data privacy. The FCGAN model demonstrated lower RMSE values com-pared to traditional local GAN models, indicating its effectiveness in handling distributed datasets. Specifically, the federated GAN achieved RMSE values of 0.0650, 0.0607, and 0.0582 for missing rates of 5%, 10%, and 15%, respectively, outperforming local GAN models. However, the study faced challenges related to the computational demands of training GANs within an FL context, as well as the complexity of the model's training process.

### 3.5 Hybrid Approaches of Federated Learning in Air Quality Monitoring

Some studies used innovative approaches by integrating FL with other methodologies or utilizing unique datasets to enhance air quality monitoring and prediction accuracy. Wardana et al. [20] explored three collaborative learning strategies—FedAvg, Clustered Model Exchange (ClustME), and Spatiotemporal Data Exchange (SpaTemp)—using edge computing devices like Raspberry Pi and Jetson Nano to predict air pollution levels. The study demonstrated that SpaTemp, which utilized spatiotemporal correlations, offered the highest prediction accuracy, with RMSE improvements ranging from 0.525% to 8.934% compared to models trained solely on local data. However, this method also required longer training times and incurred significant communication costs, posing challenges for scalability and real-time applications.

Hart and Doyle [21] proposed RealTimeAir, a federated crowd sensing system that used consumer-grade mobile sensors to provide hyper-local air quality data. The system integrated data from mobile sensors with government reference sensors to create a real-time air quality map, demonstrating the potential of FL in enhancing localized monitoring. The study achieved varying levels of correlation between mobile and government sensors, with volatile organic compounds (VOC) measurements showing the highest correlation (67.5%), while PM10 and PM2.5 showed lower correlations at 2.5% and 7.5%, respectively. Despite the innovative approach, the study identified challenges related to the accuracy of mobile sensors, particularly for pollutants like PM10 and PM2.5, and the influence of environmental factors on sensor performance.

Nguyen and Zettsu [22] applied FL to air pollution prediction using Convolutional Recurrent Neural Networks (CRNNs) in Japan's Kanto region. The study focused on managing spatially-distributed data while predicting oxidant warning levels. The model effectively captured spatial and temporal features, with the FL framework facilitating cooperative training among different regions. The study highlighted improvements in prediction accuracy, particularly for rank 3 oxidant predictions, with the introduction of new participants and the application of transfer learning within the FL framework. However, the study also faced challenges related to communication overheads, potential model attacks, and the need for secure communication protocols to ensure data privacy during model updates.

In summary, the studies reviewed demonstrate that FL holds significant promise for improving air quality monitoring by enabling decentralized data processing while maintaining privacy. The findings indicate that FL can effectively enhance the accuracy of pollutant predictions, AQI assessments, and carbon emission forecasts.

The comprehensive review of these papers is summarized in table 1.

**Table 1.** Summary of FL applications in air quality management

| Ref. | Environment | Parameters | Dataset | Methodology | Algorithm | Results | Key findings | Limitations |
|---|---|---|---|---|---|---|---|---|
| 12 | Urban | PM2.5 | 133 monitoring sites in Jiangsu Province, China (2018-2021) | Federated deep learning with ECS and adaptive gating fusion | Deep Learning (DL), LSTM, CNN | MAE: 12.38 RMSE: 19.19 R²: 0.972 | Improved accuracy and efficiency in PM2.5 forecasting using FL | Lacked discussion on security vulnerabilities; significant communication overhead requires robust network infrastructure for model updates |
| 13 | Urban | PM2.5, PM10, $SO_2$, $NO_2$, $O_3$, CO | 104,880 records from 12 cities in Fenhe River and Weihe River Plains, China (2020) | Secure FL with optimized LSTM | DL, LSTM, SSA | RMSE: 10.59 MAE: 6.26 R²: 92.93% | Improved prediction accuracy and efficiency; enhanced data privacy | Additional computational overhead; general applicability not tested; potential security vulnerabilities |
| 14 | Urban, Industrial | PM2.5, CO, NOx, Temperature, Humidity | Mumbai (Kurla, Bandra-Kurla, Nerul, Sector-19a-Nerul), (2018-2022) | Integration of CNN and LSTM within a FL framework | CNN-LSTM (DL), SVR, GRU, BiLSTM | MAE: 0.466 RMSE: 0.522 R2: 0.9877 | CNN-LSTM outperformed other models in predicting PM2.5 | Limited to Mumbai. Computational cost and energy efficiency on edge devices not discussed. |
| 15 | Urban | PM2.5, PM10, NO, $NO_2$, NOx, $NH_3$, CO, $SO_2$ | Central Pollution Control Board of Delhi (2015-2020), India | FL with UAV swarm using LSTM | DL, LSTM | RMSE: 56.222 MAE: 41.219 MAPE: 24.184 | LSTM model outperformed traditional ML models in AQI prediction; FL ensured data privacy while reducing network latency and energy consumption | Frequent sensor recalibration; UAV connectivity issues; High setup complexity |
| 16 | Urban | PM2.5, PM10, $NO_2$, CO, $O_3$ | Hourly and daily air pollutant data from Indian smart cities, expanded with noise. Source: Kaggle | FL with BGRU | DL, BGRU | MSE: 40.129 MAE: 36.659 | FL-based BGRU model achieved lower MSE and MAE compared to SVM, KNN, and RF models, demonstrating superior accuracy in predicting air quality. | Artificially expanded dataset may not fully represent real-world conditions; limited scalability testing; computationally intensive; |
| 17 | Urban | PM2.5, PM10, $SO_2$, $NO_2$, CO, $O_3$ | UCI Machine Learning Repository, Beijing, China (2013-2017) | Nested LSTM within a Multi-task Multi-channel framework and FL. | MTMC-NLSTM | MAE: 0.88 (PM2.5), 2.98 (PM10), 0.23 ($SO_2$), 0.68 ($NO_2$), 7.99 (CO), 1.51 ($O_3$) RMSE: 1.17 (PM2.5), 3.10 (PM10), 0.30 ($SO_2$), 0.74 ($NO_2$), 9.82 (CO), 2.39 ($O_3$) R²: 0.99 (PM2.5, PM10, $NO_2$, CO, $O_3$), 1.00 ($SO_2$) | The MTMC-NLSTM model achieved higher accuracy and lower error rates compared to conventional ML and DL models. Incorporating DSWT improved prediction performance by stabilizing volatile AQI data. | Limited to Beijing; scalability and real-time application not fully addressed. Increased computational complexity due to the nested LSTM structure and multi-channel approach. |

**Table 1.** Summary of FL applications in air quality management (Continued)

| Ref. | Environment | Parameters | Dataset | Methodology | Algorithm | Results | Key findings | Limitations |
|---|---|---|---|---|---|---|---|---|
| 18 | Industrial | C | Time-series data from the electricity sector in 13 countries, (01/01/2019 - 31/08/2022) Data source: China Carbon Accounting Database (CEADs). | Used SARIMA for clustering clients and FL with BiLSTM; involved data preprocessing, local SARIMA fitting, clustering, and FedAvg algorithm for model aggregation. | DT, LSTM, DNN, FedAvg, SARIMA for clustering, Federated BiLSTM for prediction | Improved MAE by 63.32% and MSE by 79.27% in optimal clusters; convergence speed improved by 73.17%. | Clustering clients based on SARIMA parameters significantly enhances the efficiency and accuracy of FL for carbon emission prediction. The model can predict carbon emissions faster and more accurately, protecting the data privacy and security of each participant. | Limited prediction accuracy due to external factors (industrial structure changes, economic growth) not being considered; uneven benefit distribution among clients; lack of incentive mechanisms for participation; exclusion of clients with unique parameter sets, and limited generalizability to other sectors beyond the electricity sector |
| 19 | Urban | $PM_{2.5}$, $PM_{10}$, $NO_x$, $O_3$, $SO_2$, CO | AQM stations of Changzhou, China (2016) | Conditional GANs within a FL framework, enhanced by Wasserstein distance and a 'Hint mask' trick | Federated Conditional GAN | RMSE: Federated GAN (0.0554 to 0.0650) vs Local GAN (0.0562 to 0.0659) | Federated GAN improved imputation accuracy, particularly in high missing rate scenarios and with non-IID data. The federated model also showed more stable training with smoother convergence. | Did not leverage time continuity; complexity in tuning hyperparameters; high computational demands; limited generalizability |
| 20 | Urban | $PM_{2.5}$, $PM_{10}$, $SO_2$, CO, $NO_2$, $O_3$ | Beijing air quality dataset from UCI Machine Learning Repository, China | Collaborative learning with edge devices using MQTT protocol | FedAvg, ClustME, SpaTemp (DL) | SpaTemp reduced RMSE by 8.934%, FedAvg by 0.588%, and ClustME by 0.725% on average compared to local training | SpaTemp demonstrated the highest accuracy in air pollution prediction by utilizing spatiotemporal data. ClustME and FedAvg also improved accuracy but were less effective than SpaTemp. | Longer training times for SpaTemp, constrained computational capabilities of edge devices, significant communication overhead, scalability concerns in larger networks |
| 21 | Urban | $PM_{10}$, $PM_{2.5}$, VOC, $NO_2$, | 930,000 data points from mobile sensors (Atmotube Pro and Flow2) and 47 hours with reference sensor in London | Federated crowd sensing using mobile sensors and government sensors | None (focus on sensor data collection and correlation) | Strong correlation for VOC (67.5%); weak correlation for $PM_{10}$ (60.8%) and $PM_{2.5}$ (74.2%) | Demonstrated feasibility of using consumer-grade sensors for hyper-local air quality monitoring; Significant spatiotemporal variations in pollutant levels identified; Developed a low pollutant exposure route finder | Low correlation between mobile sensors and reference sensors; environmental factors such as humidity affecting sensor accuracy; Limited to areas with generally low pollutant levels; Continuous dynamic calibration proposed but not implemented |
| 22 | Urban, Suburban | $SO_2$, $NO_x$, NO, $NO_2$, CO, $O_x$, NMHC, $CH_4$, THC, STM, $PM_{2.5}$ | From AEROS Kanto Region, Japan (2018-2021), | FL with local CRNN models trained on environmental sensor data; global model aggregation. | DL, CRNNs | No empirical results | Higher accuracy in predicting oxidant warning levels compared to centralized models; faster convergence for new participants through transfer learning; effective management of spatially-distributed data; efficient and scalable | High computational complexity; requires robust infrastructure and secure data protocols; communication overheads; potential model attacks |

## 4      Limitations and Future Work

Despite the promising results of FL in air quality monitoring, several limitations have been identified across the studies, alongside suggestions for future research.

A common limitation is the communication overhead and infrastructure requirements. Some studies pointed out that the frequent exchange of model updates between local devices and central servers can lead to significant communication costs, especially in large networks [12, 20]. Future work should focus on optimizing communication protocols and minimizing the frequency of updates to reduce network strain.

Scalability and generalizability are also critical issues. Huang et al. [13] and Abimannan et al. [14] demonstrated that while their models performed well in specific regions, their effectiveness in different geographic areas or under varied environmental conditions remains untested. Expanding these studies to include diverse datasets and locations will be essential for validating the broader applicability of these models.

The computational complexity of FL models, particularly those using advanced architectures like NLSTM [17] or GANs for data imputation [19], presents another challenge. These models often require significant computational resources, which can limit their deployment on resource-constrained devices. Future research should explore simplifying model architectures or developing more efficient training algorithms to mitigate these demands.

Security vulnerabilities are an ongoing concern in FL. Even though FL is designed to enhance privacy by keeping data local, studies by Huang et al. [13] and Nguyen and Zettsu [22] highlighted risks such as adversarial attacks and the potential for data reconstruction from model parameters. Strengthening encryption techniques and implementing robust security protocols will be critical areas for future work.

The accuracy of consumer-grade sensors used in studies like Hart and Doyle's [21] RealTimeAir system also poses a limitation. The variability in sensor readings, influenced by factors such as calibration and environmental conditions, can lead to inconsistencies in data quality. Future efforts should focus on improving sensor accuracy and developing standardized calibration methods to enhance data reliability.

Zhou et al. [19] also pointed out the challenges of using GANs for data imputation within an FL framework, particularly in terms of computational demands and training stability. Future research could investigate alternative, less resource-intensive imputation techniques or refine GAN training processes to improve efficiency and reliability.

Finally, the sector-specific focus studies, such as the work by Cui et al. [18] on carbon emissions in the electricity sector, limits the broader applicability of their findings. Future research should aim to apply and validate FL methodologies across various industries and develop incentive mechanisms to ensure that all participants in an FL framework benefit equitably.

Addressing these limitations through targeted research on communication efficiency, model scalability, computational optimization, security enhancement, sensor reliability, and broad applicability will be crucial for fully harnessing the potential of FL in environmental monitoring and public health protection.

## 5 Conclusion

This paper has provided a comprehensive review of FL's applications in predicting air pollutants, managing environmental data, and enhancing the accuracy and reliability of air quality forecasts.

Despite its potential, FL faces several challenges when applied to environmental monitoring. Key limitations include communication overhead, infrastructure requirements, and issues related to the generalizability of models across different regions and data distributions. Additionally, the computational complexity associated with training sophisticated models in a federated setting can hinder the widespread adoption of FL, particularly in resource-constrained environments. Moreover, FL is not without security vulnerabilities; risks such as adversarial attacks, data reconstruction from model updates, and the potential for breaches during communication between clients and servers pose significant challenges to the integrity and confidentiality of the data.

To fully realize the benefits of FL in this domain, future research must focus on optimizing FL frameworks to reduce communication and computational costs, improve scalability, and enhance model generalizability. Addressing security vulnerabilities through the development of robust encryption techniques, secure communication protocols, and methods to mitigate adversarial attacks will also be crucial. Additionally, developing infrastructure that supports the decentralized nature of FL while ensuring data security and privacy is essential.

In conclusion, while FL offers a promising path forward for air quality monitoring and environmental data management, addressing its current limitations and security vulnerabilities through targeted research and development will be essential for its successful implementation in real-world scenarios. The insights provided by this paper aim to guide future work in advancing FL as a powerful tool for improving environmental monitoring and public health.